\documentclass[journal=jctcce,manuscript=article]{achemso}
\usepackage{epstopdf}
\usepackage{amsmath}
\usepackage{lscape}
\usepackage[T1]{fontenc}

\newcommand*{\citen}[1]{%
  \begingroup
    \romannumeral-`\x % remove space at the beginning of \setcitestyle
    \setcitestyle{numbers}%
    [\cite{#1}]%
  \endgroup   
}

\author{\sc Micha\l\ Lesiuk}
\email{e-mail: lesiuk@tiger.chem.uw.edu.pl}
\author{\sc Micha\l\ Przybytek}
\affiliation{\sl Faculty of Chemistry, University of Warsaw, Pasteura 1, 02-093 Warsaw, Poland}
\author{\sc Justyna G. Balcerzak}
\affiliation{\sl Faculty of Chemistry, University of Warsaw, Pasteura 1, 02-093 Warsaw, Poland}
\author{\sc Monika Musia\l}
\affiliation{\sl Institute of Chemistry, University of Silesia, Szkolna 9, 40-006 Katowice, Poland}
\author{\sc Robert Moszynski}
\affiliation{\sl Faculty of Chemistry, University of Warsaw, Pasteura 1, 02-093 Warsaw, Poland}
\date{\today}

\title{\emph{Ab initio} potential energy curve for the ground state of beryllium dimer}

\begin{document}

\begin{abstract}
This work concerns \emph{ab initio} calculations of the complete potential energy curve and spectroscopic constants
for the ground state $X^1\Sigma_g^+$ of the beryllium dimer, Be$_2$. High accuracy and reliability of the results
is one of the primary goals of the paper. To this end we apply large basis sets of Slater-type orbitals
combined with
high-level electronic structure methods including triple and quadruple excitations. The effects of the relativity are
also
fully accounted for in the theoretical description. For the first time the leading-order
quantum electrodynamics effects are fully incorporated for a many-electron molecule. Influence of the finite nuclear mass corrections
(post-Born-Oppenheimer effects) turns out to be completely negligible for this system. The predicted
well-depth ($D_e=934.5\pm2.5\,\mbox{cm}^{-1}$) and the dissociation energy ($D_0=808.0\,\mbox{cm}^{-1}$) are in a very
good agreement with the most recent experimental data. We confirm the existence of the weakly bound twelfth
vibrational
level [Patkowski et al., Science \textbf{326}, 1382 (2009)] and predict that it lies just about 0.5 $\mbox{cm}^{-1}$
below the onset of the continuum.
\end{abstract}

\maketitle

\section{Introduction}
\label{sec:intro}

In the past decades beryllium dimer has been the subject of many studies, both experimental and
theoretical. The first calculations predicted the interaction between two closed-shell beryllium atoms to be purely repulsive
\cite{fraga61}, even when the electron correlation effects were partially included \cite{bender67}. However, more
sophisticated quantum chemistry methods became available in the late 70' and early 80' allowing to re-evaluate the
scientific consensus about the nature of the bonding in the beryllium dimer \cite{dykstra76,blomberg78,chiles81}. It was
predicted that this molecule is bound, albeit weakly, with some similarities to the noble gas dimers.

Further improvements in the theoretical description of the beryllium dimer were presented by Liu and McLean
\cite{liu80}, and somewhat later by Harrison and Handy \cite{harrison83}. Both studies reported that the single and
double
excitations with respect to the single reference wavefunction are not sufficient to describe the bonding correctly.
Inclusion of triple and quadruple excitations (either by means of full CI or multireference methods
\cite{mukherjee77,jeziorski81,mahapatra98}) is
necessary to obtain more quantitative results. This allowed to revise the bonding energy up to several hundreds of
cm$^{-1}$. Moreover, it was shown that the pathological behaviour of this system is largely due
to the near-degeneracy of the $2s$ and $2p$ energy levels of the beryllium atom. These conclusions have been confirmed
by several
other authors \cite{bausch92,fusti96a,fusti96b,starck96,keledin99,schmidt10,mitin11,khatib14}.

The fact that the beryllium dimer is an apparently simple yet challenging system has made it 
a frequent subject of state-of-the-art computational studies. At present the consensus is that the binding energy of the
beryllium dimer is in the
range 920$-$940 cm$^{-1}$ and the bond length is approximately 2.44 \AA{}. The reported values differ depending on the
employed level
of theory but it appears that the most reliable theoretical results to date were given by Martin \cite{martin99}
(944$\,\pm\,$25
cm$^{-1}$), R\o{}ggen and Veseth \cite{roeggen05} (945$\,\pm\,$15 cm$^{-1}$), Patkowski et al. \cite{patkowski07}
(938$\,\pm\,$15 cm$^{-1}$), Koput \cite{koput11} (935$\,\pm\,$10 cm$^{-1}$), and the present authors \cite{lesiuk15}
(929$\,\pm\,$1.9 cm$^{-1}$). Other notable papers are Refs. \citen{gdanitz99,pecul00,sharma14,blunt15,deible15,magoulas18,karton18} and
a
more detailed older bibliography is found in Refs. \citen{roeggen05,patkowski07}. Semiempirical (or morphed) potentials
have also been constructed for this system \cite{spirko06,meshkov14}.

Experimental studies of the beryllium dimer ground state also have a long history. The first
experimental works of Bondybey et al. \cite{bondybey84a,bondybey84b,bondybey85} were conducted in the middle 80' and
only a few
vibrational levels were observed. These incomplete data and a lacking theoretical model led to a considerably
underestimated value for the well-depth, 790$\,\pm\,$30 cm$^{-1}$. This prediction was later revised by Spirko
\cite{spirko06}
who combined the experimental results of Bondybey et al. \cite{bondybey84a,bondybey84b,bondybey85} with portions of
theoretical potentials and recommended a new value of 923 cm$^{-1}$.

A refined experiment was performed in 2009 by Merritt et al. \cite{merritt09} who reported
929.7$\,\pm\,$2 cm$^{-1}$ for the well-depth. In addition, eleven vibrational levels were characterised
\cite{bernath09}. However, to extract the potential parameters (well-depth, equilibrium distance, etc.) from the
experimental results, Merritt et al. \cite{merritt09} employed a relatively simple Morse-like potential. It vanishes too
fast (i.e.
exponentially) at large internuclear distances. This deficiency was corrected by Patkowski et al. \cite{patkowski09} who
calculated a theoretical potential energy curve with the correct $R^{-6}$ long-range behaviour, where $R$ is the internuclear distance. This potential was not
accurate enough to reproduce the experimental results with the spectroscopic accuracy, but by a simple morphing
of the potential the accuracy was greatly improved. By introducing two empirical parameters they
reproduced the experimental vibrational levels to within 1.0 cm$^{-1}$, and with five parameters the error was
further reduced to about 0.1 cm$^{-1}$.

Even more interestingly, the morphed potential of Patkowski et al. \cite{patkowski09} supported an additional (i.e.
twelfth) vibrational level. This level was not originally reported in the experimental paper of Merritt et al.
\cite{merritt09} and its existence came as a surprise. Several subsequent works tried to reproduce the
observations of Patkowski et al. \cite{patkowski09} without resorting to any empirical adjustments \cite{koput11}. In
parallel, refined
direct-potential-fit analyses provided improved (albeit purely empirical) potentials \cite{meshkov14}, supporting the
findings of Ref. \citen{patkowski09}.

In this paper we expand upon our previous work \cite{lesiuk15} where the interaction energy of the beryllium dimer at
the minimum of the potential energy curve has
been determined with help of the Slater-type orbitals \cite{slater30,slater32} by using the newly developed programs
\cite{lesiuk14a,lesiuk14b,lesiuk16}. We largely extend the results reported previously \cite{lesiuk15} and calculate the
full potential energy
curve (PEC) including corrections due to the adiabatic, relativistic, and quantum electrodynamics effects. Next, we
generate analytic fits of the
interaction potentials and solve the nuclear Schr\"{o}dinger equation to obtain the vibrational energy terms. Finally,
an extensive comparison with the existing theoretical and experimental data is given.

Atomic units are used throughout the paper unless explicitly stated otherwise. We adopt the
following conversion factors and fundamental constants: $1\,a_0 = 0.529\,177$ \AA{} (Bohr radius), 
$1\,\mbox{u}=1822.888$ (unified atomic mass unit), $1\,$H=$219\,474.63$ cm$^{-1}$ (Hartree),
$\alpha$ = $1 / 137.035\,999$ (the fine structure constant). These values are in line with the recent CODATA
recommendations \cite{mohr16}. We assume that the mass of the only stable isotope of
beryllium ($^9$Be) is $m(\mbox{Be}) = 9.012\,183 \mbox{\,u}$ which is the latest experimental value \cite{meija16}. All
data presented in this paper refer to the $^9$Be isotope. We also adopt a convention that the interaction energy is
positive whenever the underlying interaction is attractive.

\section{\emph{Ab initio} calculations}
\label{sec:abinit}

\subsection{Basis sets}
\label{subsec:basis}

In this work we use basis sets composed of the canonical Slater-type orbitals \cite{slater30,slater32}
\begin{align}
\label{sto1}
\chi_{lm}(\textbf{r};\zeta) = \frac{(2\zeta)^{n+1/2}}{\sqrt{(2n)!}}\, r^l e^{-\zeta r}\,
Y_{lm}(\theta,\phi),
\end{align}
where $\zeta>0$ is a nonlinear parameter to be optimised, and $Y_{lm}$ are spherical harmonics in 
the Condon-Shortley phase convention. In our previous paper \cite{lesiuk15} the optimisation of
the STOs basis sets has been described in detail. It the present work we employ slightly modified procedures,
so let us describe the most important differences. First, instead of the conventional well-tempering of the nonlinear
parameters for a given angular momentum ($\zeta_{lk}$) we employ a more flexible formula
\begin{align}
\label{wt}
\zeta_{lk} = \alpha_l\, \beta_l^{k + \gamma_l k^2} \;\;\; \mbox{with}\;\;\; k=0,1,2,\ldots
\end{align}
where $\alpha_l$, $\beta_l$, and $\gamma_l$ are free parameters to be optimised. For a brief discussion of
advantages of this expansion see Ref. \citen{lesiuk17}.

\begin{table}[t]
\caption{Composition of the STO basis sets for the beryllium atom used in this work.}
\begin{tabular}{c|lll}
\hline
\label{basis}
$l$ & valence & core & diffuse \\
\hline\\[-1.2em]
2 & $7s2p1d$ & $1s1p$ & $2s2p1d$ \\
3 & $8s3p2d1f$ & $2s2p1d$ & $2s2p2d1f$ \\
4 & $9s4p3d2f1g$ & $2s3p2d1f$ & $2s2p2d2f1g$ \\
5 & $9s5p4d3f2g1h$ & $3s4p3d2f1g$ & $2s2p2d2f2g1h$ \\
6 & $9s6p5d4f3g2h1i$ & $3s5p4d3f2g1h$ & $2s2p2d2f2g2h1i$ \\
\hline
\end{tabular}
\end{table}

Similarly as in the previous works we divide the basis sets into the core and valence components and employ the
correlation-consistency principle \cite{dunning89} to determine the final composition of both parts. However, in
contrast to Ref. \citen{lesiuk15} an
additional set of diffuse functions is added to each basis.  Therefore, all basis sets used here are doubly augmented.
The low-exponent functions are especially beneficial for larger internuclear distances. The final composition of
all basis sets is given in Table \ref{basis}. Other details can be obtained from
the authors upon request. For brevity, the valence-only basis sets are denoted shortly wtcc-$l$ whilst the core-valence
basis sets are abbreviated tc-wtcc-$l$. In both cases, $l$ is the highest angular momentum present in the
basis set and the double augmentation is denoted with the prefix da-, e.g. da-wtcc-$l$.

Special basis sets are used further in the paper for the calculations of the relativistic and QED
effects. In this case we modify the original da-tc-wtcc-$l$ basis sets by replacing all $s$-type functions by a common
set of twelve $1s$ orbitals. This set has been obtained by minimising the Hartree-Fock energy of the beryllium atom.
Detailed compositions of the STO basis sets used in this work (exponents and quantum numbers) are given in the
supplementary material.

In Table. \ref{coratom} we present results of the FCI calculations for the beryllium atom in the da-tc-wtcc-$l$ basis
sets. The Hartree-Fock (HF) limit is reached already with the basis set $l=5$ and we did not attempt to extrapolate the
HF results. The correlation energies are extrapolated with the help of the following formula \cite{hill85}
\begin{align}
\label{extra}
E = a + b\,(l+1)^{-3},
\end{align}
where $l$ is the highest angular momentum present in the basis, and the parameters $a$, $b$ are obtained by
least-squares fitting. The best quality of the results for the beryllium atom is obtained by extrapolation from the
$l=4,5,6$ basis sets. Two-point extrapolation from $l=5,6$ also yields good results and we found it useful in estimating the extrapolation errors. The formula (\ref{extra}) with $l=4,5,6$ will be used in all subsequent molecular calculations for
the extrapolation of the correlation energies. In Table \ref{coratom} we show that the extrapolated results for the beryllium
atom differ from the reference values of Pachucki and Komasa \cite{pachucki04a} by less than 100$\,\mu$H. Moreover,
the extrapolation reduces the error of the largest basis set by a factor of five.

\begin{table}[t]
\caption{The total nonrelativistic ($E_{\mbox{\scriptsize total}}$) and correlation energies ($E_{\mbox{\scriptsize
c}}$) of the beryllium atom calculated at the FCI level of theory (see the main text for details). The limit of the
Hartree-Fock energy is $-14.573\;023$ a.u. (all digits given are accurate).}
\label{coratom}
\begin{tabular}{lrcc}
\hline
$l$ & $N$ & $E_{\mbox{\scriptsize c}}$ / mH & $E_{\mbox{\scriptsize total}}$ \\
\hline\\[-1.0em]
2 & 31  & $-$85.976 & $-$14.658 998 \\[0.6ex]
3 & 67  & $-$91.479 & $-$14.664 502 \\[0.6ex]
4 & 124 & $-$92.994 & $-$14.666 017 \\[0.6ex]
5 & 204 & $-$93.608 & $-$14.666 631 \\[0.6ex]
6 & 316 & $-$93.902 & $-$14.666 925 \\[0.6ex]
\hline\\[-1.0em]
CBS & $\infty$ & $-$94.429 & $-$14.667 452 \\
\hline\\[-1.0em]
Ref. \citen{pachucki04a} &  & $-$94.333 & $-$14.667 356 \\
\hline
\end{tabular}
\end{table}

\subsection{Four-electron (valence) contribution}
\label{subsec:val}

Within the current computational capacities the full CI (FCI) method cannot be used for eight-electron systems with
any reasonable basis set. Therefore, in the present work we rely on a composite scheme where the total interaction
energy is divided into a set of well-defined components of different magnitudes. The largest components are calculated
most accurately, i.e. employing larger basis sets or more reliable electronic structure methods. Smaller
contributions are treated at a more approximate level of theory or even completely neglected.

It is well-known that the dominant contribution to the interaction energy of the beryllium dimer comes from
the outer valence electrons. In fact, by freezing the $1s$ core orbitals of both atoms one can still recover
approximately
90\% of the total interaction energy. Unfortunately, calculation of the valence four-electron contribution is 
challenging due to the aforementioned $2s$-$2p$ near-degeneracy of the energy levels of the beryllium atom. This leads
to a significant
multireference character of the dimer. As a result, CCSD(T) (or even CCSDT) method should not be used in an accurate
calculation of the
valence four-electron contribution to the interaction energy. To get a quantitative answer one has to use either the
FCI method or some multireference CI/CC variant. In the present paper we choose the former option, mostly because of its
black-box character and no arbitrariness, e.g. in the selection of the active orbital space.

\begin{table}
\caption{Four-electron valence calculations for the beryllium dimer; $l$ is the largest angular momentum present in the
basis set; $N_b$ is the number of basis set functions, $E_{\mbox{\scriptsize int}}^{\mbox{\scriptsize HF}}$ and
$E_{\mbox{\scriptsize int}}^{\mbox{\scriptsize FCI}}$ are interaction energies calculated at the Hartree-Fock and FCI
levels of theory, respectively. Results are given for two internuclear distances, $R$. The interaction energies are
given in $\mbox{cm}^{-1}$ and the internuclear distances in bohr.}
\begin{tabular}{crcccc}
\hline
\label{fci}
$l$ & $N_b$ &
\multicolumn{2}{c}{$R=4.75$} & 
\multicolumn{2}{c}{$R=8.00$} \\
\hline\\[-1.0em]
 &      &
$E_{\mbox{\scriptsize int}}^{\mbox{\scriptsize HF}}$  &
$E_{\mbox{\scriptsize int}}^{\mbox{\scriptsize FCI}}$ &
$E_{\mbox{\scriptsize int}}^{\mbox{\scriptsize HF}}$  &
$E_{\mbox{\scriptsize int}}^{\mbox{\scriptsize FCI}}$ \\
\hline\\[-1.0em]
2 & 54  & $-$2367.5 & 270.3 & $-$124.5 & 116.2 \\[0.6ex]
3 & 110 & $-$2324.3 & 692.2 & $-$123.5 & 155.0 \\[0.6ex]
4 & 192 & $-$2320.6 & 804.8 & $-$123.3 & 167.7 \\[0.6ex]
5 & 302 & $-$2320.5 & 831.7 & $-$123.2 & 171.1 \\[0.6ex]
6 & 448 & $-$2320.4 & 842.7 & $-$123.2 & 172.0 \\[0.6ex]
\hline\\[-1.0em]
$\infty$ & $\infty$ & $-$2320.4$\,\pm\,$0.1 & 865.5 $\pm$ 2.0 & $-$123.2 $\,\pm\,$0.1 & 174.8$\,\pm\,$1.2  \\
\hline
\end{tabular}
\end{table}

Valence four-electron FCI interaction energies were calculated with the basis sets da-wtcc-$l$, $l=2-6$. This was
accomplished by using the FCI program \textsc{Hector} \cite{przybytek14} written by one of us (MP). Canonical
Hartree-Fock orbitals
generated by external programs were used as a starting point for the FCI iterations. All FCI computations
were performed utilising the $D_{2h}$ Abelian point group symmetry. The largest basis set leads to a FCI matrix of
dimension
over one billion
($10^9$). Basis set superposition error is eliminated by applying
the counterpoise correction \cite{boys70}. In Table \ref{fci} we present exemplary results of the
valence FCI calculations. To provide a broader picture
we list
these data for two interelectronic distances, $R=4.75$ located near the minimum of PEC and $R=8.0$, already
close to the asymptotic van der Waals region.

To reach the basis set limit of the calculated quantities and estimate the corresponding errors we rely on the
CBS extrapolations. The only exception is the Hartree-Fock (HF) energy. As one can see from Table \ref{fci} the HF
contribution to the interaction energy is converged to better than 0.1 $\mbox{cm}^{-1}$ already in the basis set $l=6$.
Therefore, we simply take the value obtained with $l=6$ as the HF limit. The error of this approximation is negligible
in the present context. A more complicated situation is found for the contributions coming from the correlation energy.
To extrapolate them we employ the same formula which has been demonstrated to give reliable results for an isolated
atom, Eq. (\ref{extra}). Overall, we find that the formula (\ref{extra}) fits the raw data points very well. The
extrapolated values of the interaction energy (FCI level of theory) are also listed in Table \ref{fci}. The errors are
estimated as half of the
difference between the extrapolated results from the basis sets $l=4,5,6$ and $l=5,6$.

Let us also illustrate how important the post-CCSD(T) effects are in the calculation of the valence contributions to the
interaction energy. For example, the interaction energy calculated with the frozen-core CCSD(T) method
\cite{raghavachari89} and the $l=6$ basis set is 623.9 $\mbox{cm}^{-1}$ for the $R=4.75$
and 59.6 $\mbox{cm}^{-1}$ for $R=8.0$. Comparison with the values calculated at the FCI level of theory (cf.
Table \ref{fci}) shows that CCSD(T) recovers only approx. 75\% of the total valence interaction energy for
$R=4.75$ and 90\% for $R=8.0$. These deviations cannot be attributed to the basis set incompleteness error since a very
similar picture is obtained from the CBS-extrapolated data. Therefore, the CCSD(T) method alone is not a reasonable
level of theory for the calculation of the valence contribution to the interaction energy of the beryllium dimer. 

\begin{table*}
\caption{Core-core and core-valence contributions ($E_{\mbox{\scriptsize int,core}}^{\mbox{\scriptsize X}}$) to the
interaction energy of the beryllium dimer calculated at various levels of theory ($\mbox{X}$) in the da-tc-wtcc-$l$ basis
sets (see the main text for precise definitions of all quantities); $N_b$ is the number of basis set functions.
The interaction energies are given in $\mbox{cm}^{-1}$ and the internuclear distances in bohr.}
\begin{tabular}{crcccccc}
\hline
\label{cccv}
$l$ & $N_b$ &
\multicolumn{3}{c}{$R=4.75$} & 
\multicolumn{3}{c}{$R=8.00$} \\
\hline\\[-1.0em]
 &      &
$E_{\mbox{\scriptsize int,core}}^{\mbox{\scriptsize CCSD(T)}}$  &
$\Delta E_{\mbox{\scriptsize int,core}}^{\mbox{\scriptsize T}}$ &
$\Delta E_{\mbox{\scriptsize int,core}}^{\mbox{\scriptsize (Q)}}$ &
$E_{\mbox{\scriptsize int,core}}^{\mbox{\scriptsize CCSD(T)}}$  &
$\Delta E_{\mbox{\scriptsize int,core}}^{\mbox{\scriptsize T}}$ &
$\Delta E_{\mbox{\scriptsize int,core}}^{\mbox{\scriptsize (Q)}}$ \\
\hline\\[-1.0em]
2 & 54  & 26.1 & $-$7.4 & 1.0  & $+$0.2  & $-$0.8 & $+$0.4 \\
3 & 110 & 50.8 & $-$4.6 & 0.6  & $-$1.1  & $-$0.8 & $+$0.5 \\
4 & 192 & 54.1 & $-$3.3 & ---  & $-$1.6  & --- & --- \\
5 & 302 & 54.8 & ---    & ---  & $-$1.7  & --- & --- \\
\hline\\[-1.0em]
$\infty$ & $\infty$ & 55.6$\,\pm\,$0.8 & $-$2.3$\,\pm\,$0.5 & 0.4$\,\pm\,$0.2 & 
 $-$1.9$\,\pm\,$0.2 & $-$0.8$\,\pm\,$0.4 & $+$0.6$\,\pm\,$0.3 \\
\hline
\end{tabular}
\end{table*}

\subsection{Core-core and core-valence contributions}
\label{subsec:cccv}

Let us now consider the contribution to the total interaction energy coming from the core-core and
core-valence (i.e. inner-shell) correlations, $E_{\mbox{\scriptsize int(core)}}$. It is defined as the
difference between the exact nonrelativistic Born-Oppenheimer (BO) interaction energy and the exact four-electron valence
contribution. Fortunately, calculation of this contribution is simpler in some respects than of the valence effects. The
largest contribution to $E{\mbox{\scriptsize int(core)}}$ can be obtained with
the CCSD(T) method, $E_{\mbox{\scriptsize int(core)}}^{\mbox{\scriptsize CCSD(T)}}$. The post-CCSD(T)
contributions to $E_{\mbox{\scriptsize int(core)}}$ constitute only a few percents of the exact value --- a stark
contrast to the previous case of $E_{\mbox{\scriptsize int}}^{\mbox{\scriptsize FCI}}$.

In Table \ref{cccv} we present the inner-shell contributions to the interaction energy ($E_{\mbox{\scriptsize
int,core}}^{\mbox{\scriptsize X}}$) calculated at several different levels of theory, $\mbox{X}$. In this work we
consider $\mbox{X}=$ CCSD(T), CCSDT or CCSDT(Q). For convenience, let us also define some relative quantities
\begin{align}
 &\Delta E_{\mbox{\scriptsize int,core}}^{\mbox{\scriptsize T}} = E_{\mbox{\scriptsize int(core)}}^{\mbox{\scriptsize
CCSDT}} - E_{\mbox{\scriptsize int(core)}}^{\mbox{\scriptsize CCSD(T)}}, \\
 &\Delta E_{\mbox{\scriptsize int,core}}^{\mbox{\scriptsize (Q)}} = E_{\mbox{\scriptsize int(core)}}^{\mbox{\scriptsize
CCSDT(Q)}} - E_{\mbox{\scriptsize int(core)}}^{\mbox{\scriptsize CCSDT}},
\end{align}
Calculation of the above post-CCSD(T) corrections is computationally very intensive. For example, single-point CCSDT
calculations for the dimer in the $l=4$ basis take about a month with our computational resources. The cost of the
CCSDT(Q) method is even higher which effectively prohibits the use of basis sets larger than $l=2,3$. In the case of the
CCSDT method we managed to perform calculations up to $l=4$ only for several points on the PEC,
namely $R=4.0-5.5$ a.u. This is the region where the interaction energy is the largest and the inner-shell corrections are
the most important on the relative scale. In fact, for $R=4.75$ the inner-shell contributions stand for about 8\% of
the total interaction energy in the BO approximation (cf. Table \ref{fci}). For $R=8.0$ this ratio drops to less than 2\%.

Extrapolations of the CCSD(T) results to the CBS limit are performed with the help of the formula (\ref{extra}) with
$l=4-5$. The errors are estimated as differences between the respective values calculated with the largest
basis set and the extrapolated limit. The same technique is used for the CCSDT method where the results from $l=2,3,4$
basis
sets are available. In this case the error is estimated as 20\% of the extrapolated value. For the remaining data we
extrapolate the results by using the $l=2,3$ basis sets.
Clearly, this approach is not as reliable as the other estimates employed in this work and thus we assign an
uncertainty of 50\% to the values calculated in this way. Fortunately, the $l=2,3$ extrapolations have to be performed
only for $R>5.5$ where the post-CCSD(T) contributions are relatively small. Therefore, even if the estimated limits of
the T and T(Q) contributions were wrong by 50\%, the overall quality of the results would be affected only
marginally.

Parenthetically, a typical way to estimate the post-CCSD(T) contributions is to evaluate
them is some small basis set and add this value to the final results. As seen from Table \ref{cccv}, this is not a
particularly reasonable approach in the present context. In fact, smaller basis sets (i.e. $l=2,3$) tend to grossly
overestimate the post-CCSD(T) effects, sometimes even by a factor of $3$ or so. A similar observation has been reported by
Smith et al. \cite{smith14}

The final error of the core-core and core-valence contributions to the interaction energy is obtained by 
summing squares of the errors of all constituents ($E_{\mbox{\scriptsize int,core}}^{\mbox{\scriptsize CCSD(T)}}$,
$\Delta E_{\mbox{\scriptsize int,core}}^{\mbox{\scriptsize T}}$, $\Delta E_{\mbox{\scriptsize
int,core}}^{\mbox{\scriptsize (Q)}}$) and taking the square root. According to Table \ref{cccv} this gives the total values
of $E_{\mbox{\scriptsize int(core)}}=53.7\,\pm\,0.9\,\mbox{cm}^{-1}$ and $E_{\mbox{\scriptsize
int(core)}}=-2.1\,\pm\,0.5\,\mbox{cm}^{-1}$ for $R=4.75$ and $R=8.0$, respectively.

\begin{landscape}
\begin{table}
\caption{Relativistic corrections to the interaction energy of the beryllium dimer. The column ``valence'' gives results
calculated at the valence FCI level of theory and the column ``core'' provides the inner-shell corrections (see the main
text for details). The core correction is neglected for the two-electron Darwin and Breit terms. The interaction
energies are given in $\mbox{cm}^{-1}$ and the internuclear distances in bohr.}
\begin{tabular}{ccccccccccccc}
\hline
\label{relativ}
$l$ &
\multicolumn{2}{c}{$\langle P_4 \rangle$} &
\multicolumn{2}{c}{$\langle D_1 \rangle$} &
total &
$\langle D_2 \rangle$ &
$\langle B \rangle$ & total \\
 & valence & core & valence & core & Cowan-Griffin &  &  & Breit-Pauli  \\
\hline\\[-1.0em]
 & \multicolumn{8}{c}{$R=4.75$} \\
\hline\\[-1.0em]
2 & $-$14.1 & $-$0.40 & 10.4 & 0.31 & $-$3.8 & 0.39 & $-$0.69 & $-$4.1 \\[0.6ex]
3 & $-$15.2 & $-$0.55 & 11.2 & 0.43 & $-$4.1 & 0.43 & $-$0.71 & $-$4.4 \\[0.6ex]
4 & $-$15.4 & $-$0.57 & 11.3 & 0.44 & $-$4.2 & 0.44 & $-$0.72 & $-$4.5 \\[0.6ex]
\hline\\[-1.0em]
$\infty$ & $-$15.6$\,\pm\,$0.2 & $-$0.60$\,\pm\,$0.03 & 11.5$\,\pm\,$0.2 & 0.47$\,\pm\,$0.03 & $-$4.2$\,\pm\,$0.3 &
0.49$\,\pm\,$0.05$^{\mbox{\scriptsize a}}$ & $-$0.75$\,\pm\,$0.03$^{\mbox{\scriptsize a}}$ & $-$4.5$\,\pm\,$0.3 \\
\hline\\[-1.0em]
 & \multicolumn{8}{c}{$R=8.00$} \\
\hline\\[-1.0em]
2 & $-$0.26 & $-$0.006 & 0.19 & 0.002 & $-$0.07 & 0.009 & $-$0.016 & $-$0.08 \\[0.6ex]
3 & $-$0.39 & $-$0.008 & 0.28 & 0.003 & $-$0.12 & 0.013 & $-$0.017 & $-$0.12 \\[0.6ex]
4 & $-$0.43 & $-$0.009 & 0.30 & 0.004 & $-$0.14 & 0.015 & $-$0.017 & $-$0.14 \\[0.6ex]
\hline\\[-1.0em]
$\infty$ & $-$0.45$\,\pm\,$0.02 & $-$0.01$\,\pm\,$0.001 & 0.32$\,\pm\,$0.02 & 0.005$\,\pm\,$0.001 & $-$0.14$\,\pm\,$0.03
& 0.020$\,\pm\,$0.005$^{\mbox{\scriptsize a}}$ & $-$0.018$\,\pm\,$0.001$^{\mbox{\scriptsize a}}$ & $-$0.14$\,\pm\,$0.03
\\
\hline
\end{tabular}
\begin{flushleft}\vspace{-0.20cm}
$^{\mbox{\scriptsize a}}${\small the error estimation includes the uncertainty due to the neglected core contribution }
\end{flushleft}
\end{table}
\end{landscape}

\subsection{Relativistic corrections}
\label{subsec:rel}

To meet the high accuracy requirements of this study we must incorporate in our description of the interaction
potential the subtle effects of the relativity. As long as the constituting elements are not too heavy, the leading-order
relativistic corrections to the molecular energy levels can be calculated by perturbation theory. The approach based on
the Breit-Pauli Hamiltonian \cite{bethe75} (accurate to within $\alpha^2$) is frequently used
\begin{align}
\label{breit}
E^{(2)} &= \langle P_4 \rangle + \langle D_1 \rangle + \langle D_2 \rangle + \langle B \rangle,
\end{align}
\begin{align}
\label{p4}
\langle P_4 \rangle &= -\frac{\alpha^2}{8} \langle \sum_i \nabla_i^4 \rangle,
\end{align}
\begin{align}
\label{d1}
\langle D_1 \rangle &= \frac{\pi}{2}\alpha^2\sum_a Z_a \langle \sum_i 
\delta(\textbf{r}_{ia})\rangle,
\end{align}
\begin{align}
\label{d2}
&\langle D_2 \rangle = \pi\alpha^2\langle\sum_{i>j}\delta(\textbf{r}_{ij})\rangle,
\end{align}
\begin{align}
\label{bb}
&\langle B \rangle   = \frac{\alpha^2}{2}\langle \sum_{i>j} 
\left[\frac{\nabla_i\cdot\nabla_j}{r_{ij}}
+\frac{\textbf{r}_{ij}\cdot(\textbf{r}_{ij}\cdot\nabla_j)\nabla_i}{r_{ij}^3}\right] \rangle,
\end{align}
where $i$ and $a$ denote electrons and nuclei, respectively, $r_{XY}$ denotes the interparticle distances,
and $\langle \mathcal{O}\rangle$ is the expectation value of an operator $\mathcal{O}$. 
Further in the paper the above corrections are referred shortly to as the
mass-velocity, one-electron Darwin, two-electron Darwin, and orbit-orbit terms (in the order of appearance). Moreover,
the sum of $\langle P_4 \rangle$ and $\langle D_1 \rangle$ terms is called the Cowan-Griffin correction
\cite{cowan76}, and the names ``orbit-orbit'' and ``Breit'' shall be used interchangeably for the term (\ref{bb}).

In the calculations of the relativistic effects we adopt the following approach. Similarly as for the nonrelativistic
energies, the relativistic
contributions are divided into the valence and core components. In the case of the two-electron relativistic corrections,
$\langle D_2 \rangle$ and $\langle B \rangle$, we
neglect the core contribution. This is justified because the two-electron contributions are by an order of magnitude
smaller than $\langle P_4 \rangle$ and $\langle D_1 \rangle$ terms, and the core components are further by an
order of magnitude smaller than the valence effects. This was verified by carrying out FCI calculations in small basis sets. 
We estimated that the neglected terms would bring a contribution of only about 0.01 $\mbox{cm}^{-1}$ to the interaction energy at the minimum of PEC. Thus,
they are entirely negligible in the present study, cf. Ref. \citen{lesiuk15}.
Nonetheless, we add an additional uncertainty of 5\% to the calculated two-electron relativistic effects due to the
neglected core contributions which is probably a very conservative estimation.

Extrapolations of the relativistic corrections to the complete basis set limit are performed with help of Eq.
(\ref{extra}). The only exception is the two-electron Darwin term where the $l^{-1}$ convergence pattern is found. This
is consistent with the numerical experience of Refs. \citen{ottschofski97,halkier00} and theoretical findings of
Kutzelnigg \cite{kutz08}. In all cases the errors are estimated as the difference between the extrapolated result and the
value in the largest basis set.

The valence relativistic corrections are evaluated with the help of the FCI method. The
core corrections to the $\langle P_4 \rangle$ and $\langle D_1 \rangle$ terms were computed at the CCSD(T) level of
theory. In Table \ref{relativ} we show a short summary of the results for two interatomic distances. One can see
that in both cases the relativistic contribution to the interaction energy is
non-negligible. Close to the minimum of PEC the relativistic effects decrease the interaction
energy by about $5\,\mbox{cm}^{-1}$ (or 0.5\%) --- a surprisingly large amount for a system as light as the beryllium
dimer.

\subsection{Other corrections}
\label{subsec:qedad}

\begin{table*}
\caption{Quantum electrodynamics contributions to the interaction energy of the beryllium dimer. The core corrections
are neglected. The interaction energies are given in $\mbox{cm}^{-1}$ and the internuclear distances in bohr.}
\begin{tabular}{ccccccccccccc}
\hline
\label{qedt}
$l$ &
$E_1^{(3)}$ &
$E_2^{(3)}$ &
$\langle H_{AS} \rangle$ & total QED \\
\hline\\[-1.0em]
 & \multicolumn{4}{c}{$R=4.75$} \\
\hline\\[-1.0em]
2 & 0.31 & $-$0.011 & $-$0.012 & 0.29 \\[0.6ex]
3 & 0.34 & $-$0.012 & $-$0.013 & 0.32 \\[0.6ex]
4 & 0.34 & $-$0.012 & $-$0.014 & 0.32 \\[0.6ex]
\hline\\[-1.0em]
$\infty$ & 0.35$\,\pm\,$0.01 & $-$0.014$\,\pm\,$0.002 & $-$0.020$\,\pm\,$0.007 & 0.32$\,\pm\,$0.02 \\
\hline\\[-1.0em]
 & \multicolumn{4}{c}{$R=8.00$} \\
\hline\\[-1.0em]
2 & $-$0.006 & 0.0$^{\mbox{\scriptsize a}}$ & 0.0$^{\mbox{\scriptsize a}}$ & $-$0.006 \\[0.6ex]
3 & $-$0.008 & 0.0$^{\mbox{\scriptsize a}}$ & 0.0$^{\mbox{\scriptsize a}}$ & $-$0.008 \\[0.6ex]
4 & $-$0.009 & 0.0$^{\mbox{\scriptsize a}}$ & 0.0$^{\mbox{\scriptsize a}}$ & $-$0.009 \\[0.6ex]
\hline\\[-1.0em]
$\infty$ & $-$0.009$\,\pm\,$0.001 & 0.0$^{\mbox{\scriptsize a}}$ & 0.0$^{\mbox{\scriptsize a}}$ & 
$-$0.009$\,\pm\,$0.001 \\
\hline
\end{tabular}
\begin{flushleft}\vspace{-0.20cm}
$^{\mbox{\scriptsize a}}${\small below $10^{-3}$ $\mbox{cm}^{-1}$; impossible to calculate reliably due to large
cancellations between the dimer and the monomers }
\end{flushleft}
\end{table*}

Let us now move to the calculation of the quantum electrodynamics (QED) effects. According to the so-called
nonrelativistic QED theory
the leading-order post-Breit-Pauli correction to the energy of a molecule in the singlet
spin state reads \cite{caswell86,pachucki93,pachucki98,pachucki04a,pachucki05}
\begin{align}
E^{(3)} &= E_1^{(3)} + E_2^{(3)} + \langle H_{AS} \rangle,
\end{align}
where $E_1^{(3)}$ and $E_2^{(3)}$ are the one- and two-electron contributions
\begin{align}
\label{qed1}
 E_1^{(3)} &= \frac{8\alpha}{3\pi}\left(\frac{19}{30}-2\ln \alpha - \ln k_0\right)\langle D_1 \rangle,\\
 E_2^{(3)} &= \frac{\alpha}{\pi} \left(\frac{164}{15}+\frac{14}{3}\ln \alpha\right)\langle D_2 \rangle,
\end{align}
and $\langle H_{AS} \rangle$ is the Araki-Sucher correction \cite{araki57,sucher58} given by the formula
\begin{align}
\label{ara}
\langle H_{AS} \rangle = -\frac{7\alpha^3}{6\pi} \langle \sum_{i>j} 
P\left(r_{ij}^{-3}\right)\rangle,
\end{align}
where $P\left(r_{ij}^{-3}\right)$ denotes the regularised $r_{ij}^{-3}$ distribution,
\begin{align}
\begin{split}
\langle P\left(r_{ij}^{-3}\right)\rangle &= \lim_{a\rightarrow 0}\langle
\theta(r_{ij}-a)\,r_{ij}^{-3}
+4\pi\,(\gamma_E+\ln a)\,\delta(\textbf{r}_{ij}) \rangle, 
\end{split}
\end{align}
and $\gamma_E\approx0.57722\ldots$ is the Euler-Mascheroni constant. The other new quantity 
appearing in the above
expressions is the Bethe logarithm \cite{bethe75,schwartz61}, $\ln k_0$.

Let us note that the Araki-Sucher term is formally a two-electron expectation value so it could have been included in
$E_2^{(3)}$. However, we prefer to consider it separately due to its different nature. Additionally, it may be slightly
confusing that the name ``one-electron correction`` is assigned to $E_1^{(3)}$ as $\ln k_0$ is a many-electron quantity.
However, this establishes a close parallel between the QED and relativistic corrections, cf. Eqs.
(\ref{breit})-(\ref{bb}).

Calculation of the complete leading-order QED corrections for many-electron molecules is notoriously
difficult. This is due to the presence of two complicated terms: $\langle H_{AS} \rangle$ and $\ln k_0$. A general method
to evaluate the Araki-Sucher correction with the help of the standard quantum chemistry methods has been presented only
very recently \cite{balcerzak17}. This approach has been used in the present paper. Similarly as for the two-electron
relativistic corrections we neglect the core contributions to the $\langle H_{AS} \rangle$ term.

Even more complicated issue is evaluation of the Bethe logarithm, $\ln k_0$. Fortunately, for all molecules where the
Bethe logarithm is known accurately (hydrogen molecular ion \cite{bukowski92,korobov04,korobov06,korobov13}, hydrogen
molecule \cite{kpisz09,komasa11}, helium dimer \cite{przybytek10,cencek12}) $\ln k_0$ depends weakly on
the internuclear distance, $R$. Therefore, as long as one is not interested in the interaction potential for a very
small $R$, the atomic value of $\ln k_0$ can be adopted. The Bethe logarithm for the beryllium atom has been evaluated
recently by Pachucki and Komasa \cite{pachucki13} and we adopt their value, $\ln k_0 = 5.75034$.

With the help of this approximation $E_1^{(3)}$ and $E_2^{(3)}$ are obtained by scaling the $\langle D_1
\rangle$ and $\langle D_2 \rangle$ corrections. The scaling factors do not depend on $R$ and in the present case are
approximately equal to $0.0293$ and $-0.0279$ for $E_1^{(3)}$ and $E_2^{(3)}$, respectively. In Table \ref{qedt} we
present the values of all QED corrections for $R=4.75$ and $R=8.00$. They were calculated with the same basis
sets as the relativistic effects. The total QED correction is only by an order of magnitude
smaller than the Breit-Pauli contribution. This is somewhat contradictory to the estimates based on the order in
$\alpha$, but a similar situation is found, e.g. for the hydrogen molecule \cite{kpisz09}. There are
known examples where the QED corrections are even larger than the relativistic ones \cite{puchalski15,pachucki06a}.
Fortunately,
such anomalies are absent in the higher-order QED effects\cite{puchalski16}.

The one-electron term $E_1^{(3)}$ dominates the total QED correction and the two-electron effects are smaller by a
factor of 20$-$30.
Interestingly, the total QED contribution (\ref{qed1}) increases the interaction energy of the beryllium dimer at every
point of PEC (i.e. it is attractive). Unfortunately, for larger $R$ we have encountered significant
difficulties in
calculation of the two-electron QED effects. This is mostly due to the fact that they are very small ($<10^{-3}$
$\mbox{cm}^{-1}$) and subtraction between the dimer and monomer values leads to a large cancellation of significant
digits (cf. Table \ref{qedt}). Therefore, further in the text we neglect the two-electron QED effects and include only
the $E_1^{(3)}$ term as the dominant contribution to the interaction energy.

We can also estimate the influence of the higher-order relativistic and QED effects on the total interaction energy of
the beryllium
dimer. Experiences for the helium atom \cite{pachucki06a,pachucki06b} and the hydrogen molecule \cite{puchalski16}
suggest that the dominant term of the $\alpha^4$ QED correction is the so-called one-loop diagram \cite{eides01} given by the following formula
\begin{align}
E^{(4)}_{\rm one-loop} = 16 \alpha^2 \left( \frac{427}{192}-\ln 2\right)\langle D_1 \rangle,
\end{align}
for a molecule in the singlet electronic state. Since the one-electron Darwin term $\langle D_1 \rangle$ has already been
calculated in the course of this work, the one-loop term is straightforward to obtain. We find that it contributes as
little as about $0.02\,\mbox{cm}^{-1}$ near the minimum of PEC. Therefore, the higher-order QED effects can safely be 
neglected within the present accuracy standards and one can rest assured that the QED perturbative series is
sufficiently well-converged already in the third-order.

Finally, we consider the finite nuclear mass effects. As it is well-known, the leading-order finite nuclear mass
correction to the energy of a molecule is the so-called adiabatic correction (also known as the diagonal
Born-Oppenheimer correction). As indicated in several previous works \cite{koput11,lesiuk15}, this correction is rather
small in the present
case and we calculate it at the CCSD level of theory \cite{gauss06}. For this purpose we employ the Gaussian-type
orbitals (GTOs) basis developed by Prascher et al \cite{prascher11}. Note that this is the only element of our
calculations where we resort to GTOs. Our results indicate that the adiabatic effects are indeed very small for
the ground state of the beryllium dimer. For example, they amount only to $-0.14\,\mbox{cm}^{-1}$ and
$-0.02\,\mbox{cm}^{-1}$ for $R=4.75$ and $R=8.0$, respectively, in line with the simplistic estimates based on
the scaling of the BO interaction energy by the atomic mass. This justifies the neglect of the post-Born-Oppenheimer
effects in this study.

\begin{table}[b!]
\caption{Optimised parameters of the fit (\ref{ffit}) for the non-relativistic Born-Oppenheimer potential energy curve
[$V^{\mbox{\scriptsize BO}}(R)$]. The symbol $X[\pm n]$ stands for $X\cdot 10^{ \pm  n}$.}
\label{bofit}
\begin{tabular}{clcl}
\hline
parameter & \multicolumn{1}{c}{$V^{\mbox{\scriptsize BO}}(R)$} &
parameter & \multicolumn{1}{c}{$V^{\mbox{\scriptsize BO}}(R)$} \\[0.4ex]
\hline\\[-2.1ex]
$a$       & $+$6.453\,[$-$01] & 
$C_6$     & $+$2.140\,[$+$02]$^{\mbox{\scriptsize a}}$ \\[0.6ex]
$b$       & $+$5.123\,[$-$02] & 
$C_8$     & $+$1.023\,[$+$04]$^{\mbox{\scriptsize a}}$ \\[0.6ex]
$\eta$    & $+$6.052\,[$-$01] & 
$C_{10}$  & $+$5.165\,[$+$05]$^{\mbox{\scriptsize a}}$ \\[0.6ex]
\hline\\[-2.1ex]
$c_{0}$   & $-$7.063\,[$+$05] & $c_{9}$    & $+$1.501\,[$+$02] \\[0.6ex]
$c_{1}$   & $+$1.647\,[$+$06] & $c_{10}$   & $-$1.359\,[$+$01] \\[0.6ex]
$c_{2}$   & $-$1.778\,[$+$06] & $c_{11}$   & $+$9.449\,[$-$01] \\[0.6ex]
$c_{3}$   & $+$1.179\,[$+$06] & $c_{12}$   & $-$4.950\,[$-$02] \\[0.6ex]
$c_{4}$   & $-$5.376\,[$+$05] & $c_{13}$   & $+$1.888\,[$-$03] \\[0.6ex]
$c_{5}$   & $+$1.786\,[$+$05] & $c_{14}$   & $-$4.944\,[$-$05] \\[0.6ex]
$c_{6}$   & $-$4.473\,[$+$04] & $c_{15}$   & $+$7.944\,[$-$07] \\[0.6ex]
$c_{7}$   & $-$8.612\,[$+$03] & $c_{16}$   & $-$5.899\,[$-$09] \\[0.6ex]
$c_{8}$   & $-$1.288\,[$+$03] & \\[0.6ex]
\hline
\end{tabular}
\begin{flushleft}\vspace{-0.2cm}
$^{\mbox{\scriptsize a}}${\small fixed - taken from Ref. \citen{porsev06} }\;
\end{flushleft}
\end{table}

\begin{figure}[t!]
\includegraphics[scale=1.0]{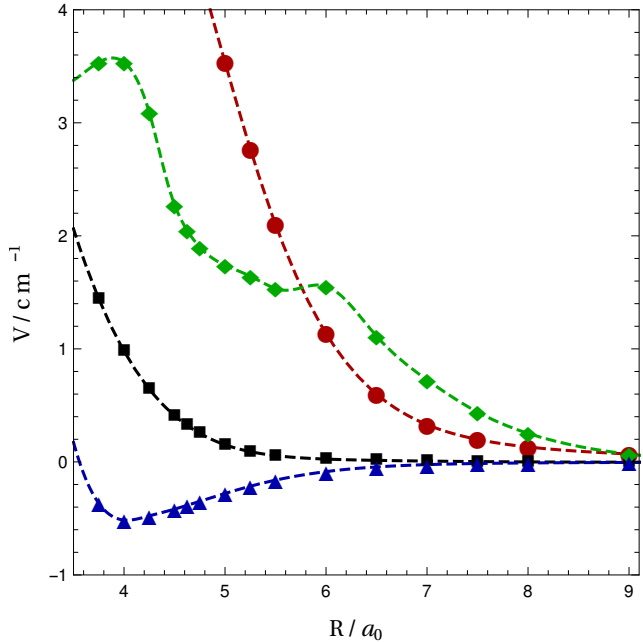}
\caption{Minor corrections to PEC of the beryllium dimer as a function of the internuclear
distance. By red dots, black squares, green diamonds, and blue triangles we denote, respectively, the one-electron relativistic correction,
two-electron relativistic correction, post-CCSD(T) inner-shell correction, and one-electron QED
correction.}
\label{figtiny}
\end{figure}

The influence of various minor physical effects on the interaction energy of the beryllium dimer is illustrated in
Fig. \ref{figtiny}. By the term ``minor'' we mean all non-negligible contributions calculated in the course of this
work other than the four-electron FCI and the inner-shell CCSD(T) contributions (which together constitute about 99\% of
the total value). Overall, the one-electron relativistic corrections are the most important among the quantities
included in Fig. \ref{figtiny}, followed by the inner-shell post-CCSD(T) effects. The remaining corrections shown in 
Fig. \ref{figtiny} are almost by
an order of magnitude smaller than the latter two. An interesting feature visible in Fig. \ref{figtiny} is a pronounced
hump in the inner-shell post-CCSD(T) corrections curve. We believe that this feature is related to the change in the
character of the chemical bond as argued in Ref. \citen{khatib14}.

\subsection{Computational details}
\label{subsec:kompot}

Most of the electronic structure calculations described above were carried out with help of the \textsc{Gamess} program
suite \cite{gamess1}. The only exceptions are the FCI calculations (performed with \textsc{Hector} program
\cite{przybytek14}), higher-order coupled
cluster methods (\textsc{AcesII} \cite{aces2}) and calculations of the adiabatic correction (\textsc{CFour}
\cite{cfour}). Matrix elements of the
orbit-orbit and Araki-Sucher operators were not evaluated directly in the STOs basis, but with help of the Gaussian
fitting technique by using twelve GTOs representing a single exponential orbital.
For a single point of the curve ($R=4.75$) we recomputed all quantities employing fifteen GTOs but the changes were
marginal.

To create a complete PEC we selected the following grid of internuclear distances
\begin{align*}
 &\mbox{from $R=3.75$ to $R=5.50$ in steps of $0.25$,}\\
 &\mbox{from $R=5.50$ to $R=8.00$ in steps of $0.50$,}\\
 &\mbox{from $R=8.00$ to $R=15.0$ in steps of $1.00$,}\\
 &\mbox{from $R=15.0$ to $R=25.0$ in steps of $2.50$,}
\end{align*}
in atomic units. Additionally, we evaluated a single point at $R=4.625$ to improve the
description of the minimum of PEC. This gives a total number of 25 points with increasing
spacings, so that the grid is more dense in regions with larger variations of the total
interaction energy.

\section{Analytic fits of the potentials}
\label{sec:fits}

The raw \emph{ab initio} data points were fitted with the conventional analytic form frequently used for the atom-atom
interactions
\begin{align}
\label{ffit}
\begin{split}
 V(R) = e^{-a R -b R^2} \sum_{k=0}^{N_p} c_k R^k 
 - \sum_{n=3}^{N_a} f_{2n}(\eta R)\,\frac{C_{2n}}{R^{2n}},
\end{split}
\end{align}
where $f_{2n}(\eta R)$ are the Tang-Toennies damping functions \cite{tang84}.
This expression contains three nonlinear parameters ($a$, $b$, and $\eta$) and an adjustable number of the linear
parameters ($N_p$) and the asymptotic constants ($N_a$). In the case of the BO potential we employ the
asymptotic constants $C_{2n}$ evaluated with more accurate theoretical methods. For the $C_6$, $C_8$, and $C_{10}$
dispersion coefficients we adopt the values reported by Porsev and Derevianko \cite{porsev02,porsev06} which are in a
very good agreement with the earlier results of Mitroy and Bromley \cite{mitroy03}. No reliable data is available for
the higher-order constants so they are neglected here. 

\begin{figure}[t!]
\includegraphics[scale=1.0]{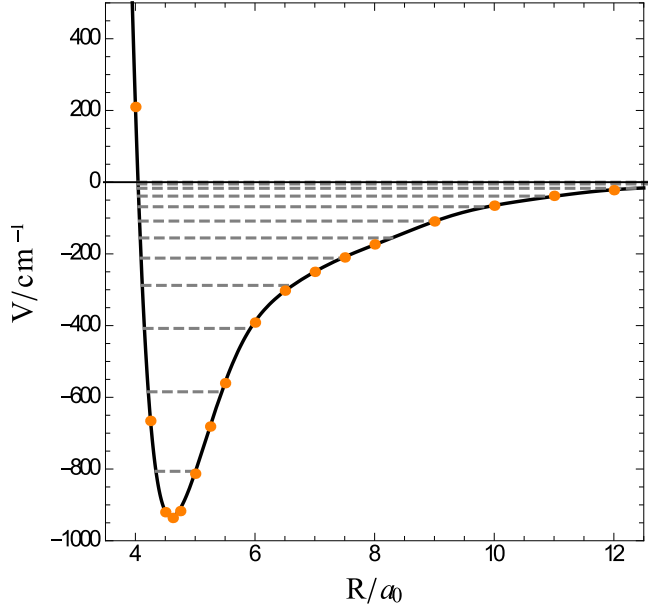} 
\caption{Complete PEC for the $X^1\Sigma_g^+$ state of Be$_2$ (solid black line);
orange dots are the extrapolated \emph{ab initio} data points. The horizontal dashed lines are energies of the $J=0$
vibrational levels. The horizontal black solid line denotes the onset of the continuum.}
\label{figkurwy}
\end{figure}

Unfortunately, the BO results for the beryllium dimer are very difficult to fit with a smooth analytic function. This
is clearly related to the unusual shape of this curve illustrated in Fig. \ref{figkurwy}. We needed
as many as 16 parameters to obtain an accurate fit of the BO potential. The optimised values are given in 
Table \ref{bofit}. Further work is necessary to reduce the number of parameters.

\begin{table}[t!]
\caption{Optimised parameters of the fit (\ref{ffit}) for the one-electron relativistic corrections, see Eqs.
(\ref{p4}) and (\ref{d1}). The symbol $X[\pm n]$ stands for $X\cdot 10^{\pm n}$.}
\label{parrel1}
\begin{tabular}{cccc}
\hline
parameter & $V^{\mbox{\scriptsize P4}}(R)$
          & $V^{\mbox{\scriptsize D1}}(R)$ \\[0.4ex]
\hline\\[-2.1ex]
$a$        & $+$1.180\,[$+$00] & $+$1.128\,[$+$00] \\[0.6ex]
$b$        & $+$1.099\,[$-$01] & $+$3.231\,[$-$02] \\[0.6ex]
$\eta$     & $+$9.410\,[$-$01] & $+$1.083\,[$+$00] \\[0.6ex]
\hline\\[-2.1ex]
$c_{0}$    & $-$1.759\,[$+$00] & $-$1.556\,[$-$01] \\[0.6ex]
$c_{1}$    & $+$9.444\,[$-$01] & $+$6.433\,[$-$02] \\[0.6ex]
$c_{2}$    & $-$1.298\,[$-$01] & $-$5.854\,[$-$03] \\[0.6ex]
\hline\\[-2.1ex]
$C_6$      & $+$4.106\,[$-$01] & $-$2.717\,[$-$01] \\[0.6ex]
$C_8$      & $+$6.406\,[$+$01] & $-$4.371\,[$+$01] \\[0.6ex]
$C_{10}$   & $-$2.908\,[$+$03] & $+$1.786\,[$+$03] \\[0.6ex]
\hline
\end{tabular}
\end{table}

\begin{table}[t!]
\caption{Optimised parameters of the fit (\ref{ffit}) for the two-electron relativistic corrections, see Eqs.
(\ref{d2}) and (\ref{bb}) for the definitions. The symbol $X[\pm n]$ stands for $X\cdot 10^{\pm n}$.}
\label{parrel2}
\begin{tabular}{cccc}
\hline
parameter & $V^{\mbox{\scriptsize D2}}(R)$
          & $V^{\mbox{\scriptsize B}}(R)$ \\[0.4ex]
\hline\\[-2.1ex]
$a$        & $+$4.278\,[$-$01] & $-$7.024\,[$-$02] \\[0.6ex]
$b$        & $+$7.388\,[$-$02] & $+$9.394\,[$-$02] \\[0.6ex]
$\eta$     & $+$1.128\,[$+$00] & $+$8.485\,[$+$00] \\[0.6ex]
\hline\\[-2.1ex]
$c_{0}$    & $-$5.700\,[$-$04] & $-$9.031\,[$-$05] \\[0.6ex]
$c_{1}$    & $+$2.373\,[$-$04] & $+$2.196\,[$-$05] \\[0.6ex]
$c_{2}$    & $-$2.343\,[$-$05] & $-$1.519\,[$-$06] \\[0.6ex]
\hline\\[-2.1ex]
$C_4$      & --- & $+$1.839\,[$-$04] \\[0.6ex]
$C_6$      & $-$1.582\,[$-$02] & $-$3.636\,[$-$03] \\[0.6ex]
$C_8$      & $-$7.303\,[$-$01] & $-$3.762\,[$-$02] \\[0.6ex]
$C_{10}$   & $-$2.015\,[$+$02] & --- \\[0.6ex]
\hline
\end{tabular}
% \begin{flushleft}\vspace{-0.2cm}
% $^{\mbox{\scriptsize a}}${\small fixed; see the main text for details }\;
% \end{flushleft}
\end{table}

The generic formula (\ref{ffit}) is also used for the fitting of the relativistic corrections. Each correction defined by
Eq. (\ref{breit}) is fitted separately. Unfortunately, we are not aware of any reliable asymptotic constants which could
be used for the present purposes. Therefore, we use Eq. (\ref{ffit}) with $N_p=2$ and $N_a=3$ and obtain approximate
dispersion coefficients directly from the fit. This leaves nine free parameters to be determined
by the fitting procedure which is sufficient to obtain a satisfactory accuracy. The only exception from the procedure
described above is found for the orbit-orbit correction,
Eq. (\ref{bb}), which possesses the $C_4/R^4$ long-range asymptotics \cite{meath66}. Therefore, instead of the
$C_6-C_{10}$
coefficients we use leading-order $C_4-C_8$ constants as free parameters. The one-electron QED correction is
obtained by scaling of the
$\langle D_1 \rangle$ relativistic correction according to the formula (\ref{qed1}). The fitting errors are by an order
of magnitude smaller than the estimated uncertainties of the respective theoretical results. The final optimised
values of the fitting parameters are given in Table \ref{parrel1} and Table \ref{parrel2} and the complete PEC is illustrated in
Fig. \ref{figkurwy}. The raw \emph{ab initio} data used for fitting are included in the supplementary material of this paper.

\section{Spectroscopic data}
\label{sec:spectro}

The total PEC generated in this work was used to calculate the spectroscopic parameters of the
ground state of the beryllium dimer. The well-depth ($D_e$) and the equilibrium bond length ($R_e$) are obtained by
finding the minimum of the PEC numerically. This gives the values of $D_e=934.5\,\mbox{cm}^{-1}$ and
$R_e=2.4425\,$\AA{}. We estimate that the error of the theoretically determined well-depth is at most
$2.5\,\mbox{cm}^{-1}$. 

Let us now compare these results with the experimental and theoretical data available in the
literature. The original experimental result of Merritt \emph{et al.} \cite{merritt09} is $D_e=929.7\,\mbox{cm}^{-1}$
employing
the expanded Morse oscillator (EMO) model of the potential. This choice is less than ideal for the beryllium dimer due
to 
an unphysical decay at large interatomic distances. This deficiency was first pointed out by Patkowski \emph{et al.}
\cite{patkowski09}. 
who employed ``morphed'' \emph{ab initio} potential energy curves with the correct asymptotics. Depending on the number
of parameters used in the morphing procedure the value
of $D_e$ varied in the range $933.0-934.6\,\mbox{cm}^{-1}$, in almost perfect agreement with the present result. A
similarly good agreement is found with the recent work of Meshkov \emph{et al.} \cite{meshkov14} where two empirical
potentials have been
determined by the direct-potential-fit procedure. The Morse-long range (MLR) and Chebyshev polynomial expansion
(CPE) functions give $D_e=934.8\pm0.3\,\mbox{cm}^{-1}$ and $D_e=935.0\pm0.3\,\mbox{cm}^{-1}$, respectively. The most
recent \emph{ab initio} result of Koput \cite{koput11}, $D_e=935\pm10\,\mbox{cm}^{-1}$, is also well within the present
error bars. In
Table \ref{compare} we show a compilation of the spectroscopic data obtained from selected
experimental measurements, semi-empirical/morphed potentials, and pure \emph{ab initio} calculations.

Let us note that in our recent theoretical work \cite{lesiuk15} we have predicted the well-depth to be
$D_e=929.0\pm1.9\,\mbox{cm}^{-1}$. This is outside the error bars of the present work and \emph{vice versa}.
Both results have been obtained with a very similar method, so this discrepancy requires a more detailed explanation. 
This difference can mostly be attributed to the fact that the internuclear distance adopted in Ref.
\citen{lesiuk15}
does not correspond to the true minimum of the theoretical PEC. In Ref. \citen{lesiuk15} the calculations were
performed only for a single $R$ from the work of Merritt et al. \cite{merritt09} ($R_e=2.4536\,$\AA{}). This value
differs from the minimum of the potential energy curve determined in the present work by more than $0.01$\AA{}. To
resolve the discrepancy we repeated the calculations of Ref. \citen{lesiuk15} using exactly the same methodology
(basis sets, extrapolations, electronic structure methods, etc.) but with the value of $R$ found here. We obtained
$D_e=931.7\,\mbox{cm}^{-1}$ which is significantly closer to the results of the present work. A slightly increased
uncertainty of the present results comes mostly from the inaccuracy of the fit. We recommend that the present result
($D_e=934.5\,\pm\,2.5\,\mbox{cm}^{-1}$) is referenced in other works instead of the value given in Ref. \citen{lesiuk15}.

\begin{table*}
\caption{Comparison of the selected empirical and theoretical results for the ground state of the
beryllium dimer. The energies are given in cm$^{-1}$ and the equilibrium distances in \AA{}ngstr\"{o}m, \AA{}.
The symbol $E_b(\nu=11)$ denotes the binding energy of the last (twelfth) vibrational level and the remaining
abbreviations are defined in the main text.}
\label{compare}
\begin{tabular}{cccccc}
\hline
Ref. & method & $D_e$ & $D_0$ & $R_e$ & $E_b(\nu=11)$ \\[0.4ex]
\hline\\[-2.5ex]
\multicolumn{6}{c}{empirical/morphed potentials} \\
\hline\\[-2.1ex]
 \citen{bondybey84b} & $\nu$ extrapolation & 790$\,\pm\,$30 & 660 & 2.45 & --- \\
 \citen{merritt09}   & EMO & 929.7$\,\pm\,$2.0 & 806.53 & 2.4536 & --- \\
 \citen{spirko06}    & morphed 3-param. & 922.9 & 795.0 & 2.4382 & --- \\
 \citen{patkowski09} & morphed 5-param. & 934.6 & 807.4 & 2.438 & 0.42 \\
 \citen{meshkov14}   & MLR potential fit & 934.8$\,\pm\,$0.3 & 808.16 & 2.445 & 0.518 \\
 \citen{meshkov14}   & CPE potential fit & 935.0$\,\pm\,$0.3 & 808.20 & 2.445 & 0.521 \\
\hline\\[-2.5ex]
\multicolumn{6}{c}{pure \emph{ab initio} potentials} \\
\hline\\[-2.1ex]
 \citen{martin99}    & CCSD(T)+FCI & 944$\,\pm\,$25 & 816 & 2.440 & --- \\
 \citen{gdanitz99}   & CAS $r_{12}$-MR-ACPF & 898$\,\pm\,$8 & 772 & 2.444 & --- \\
 \citen{roeggen05}   & EXRHF & 945$\,\pm\,$15 & 819 & 2.452 & --- \\
 \citen{patkowski07} & CCSD(T)+FCI & 938$\,\pm\,$15 & --- & 2.44 & --- \\
 \citen{koput11}     & CCSD(T)+FCI & 935$\,\pm\,$10 & 808.3 & 2.444 & 0.4 \\
 this work          & see the text & 934.5$\,\pm\,$2.5 & 808.0 & 2.4425 & 0.51 \\
\hline
\end{tabular}
\end{table*}

Finally, we solve the (radial) nuclear Schr\"{o}dinger equation with the help of the DVR method \cite{colbert92} to
obtain the vibrational energy levels. The results are listed in Table \ref{levels} and compared with the experimental
results of Merritt et al. \cite{merritt09} We find a very good agreement between the theoretical and empirical
vibrational energy terms. The average deviation is only about 1 cm$^{-1}$ indicating that the spectroscopic accuracy
has indeed been achieved. Additionally, we note that the experimental uncertainty of the data of Ref. \citen{merritt09}
is
about 0.5 cm$^{-1}$, so that the accuracy of our \emph{ab initio} results might be slightly better than the
average deviation suggests. Crucially, our PEC supports twelve vibrational energy levels confirming the prediction of
Patkowski
et al. \cite{patkowski09} The last vibrational level calculated with the current PEC lies just about 0.5 cm$^{-1}$ below
the onset of the continuum. This is in a good agreement both with Ref. \citen{patkowski09} where a value in
the range 0.40$-$0.44 cm$^{-1}$ was predicted, and with the more recent Ref. \citen{meshkov14} where the value of 0.52
cm$^{-1}$ was obtained. Despite our results favour the latter value, the accuracy of PEC developed in this work is not
sufficient to give a definite answer.

\begin{table}
\caption{Comparison of the vibrational spectra $E(\nu)-E(\nu=0)$ for the $X^1\Sigma_g^+$ state of the beryllium dimer. The
experimental values from Ref. \cite{merritt09} are listed in the second column, the \emph{ab initio} values obtained in
this work are listed in the third column, and the deviations between the latter two are given in the last column.
All values are given in cm$^{-1}$.}
\label{levels}
\begin{tabular}{cccccc}
\hline\\[-2.5ex]
$\nu$ & \multicolumn{3}{c}{$E(\nu)-E(\nu=0)$} \\[0.4ex]
\hline\\[-2.5ex]
 & exp. & this work & deviation \\[0.4ex]
\hline\\[-2.5ex]
1  & 222.6 & 223.5 & 0.9 \\
2  & 397.1 & 398.9 & 1.8 \\
3  & 518.1 & 520.5 & 2.4 \\
4  & 594.8 & 596.6 & 1.8 \\
5  & 651.5 & 652.9 & 1.4 \\
6  & 698.8 & 699.9 & 1.1 \\
7  & 737.7 & 738.5 & 0.8 \\
8  & 768.2 & 769.1 & 0.9 \\
9  & 789.9 & 790.5 & 0.6 \\
10 & 802.6 & 803.0 & 0.4 \\
11 & ---   & 807.5 & --- \\
\hline
$\delta_{rms}$ & --- & --- & 1.3 \\
\hline
\end{tabular}
\end{table}

\section{Conclusions}
\label{sec:concl}

The present work reports a detailed first-principles theoretical study of the ground electronic state of the beryllium dimer.
An
accurate \emph{ab initio} potential energy curve for this system has been calculated with a composite scheme employing
several quantum-chemical methods and large basis sets composed of Slater-type orbitals. The dominant (four-electron)
valence contribution to the interaction energy has been calculated at the FCI level of theory. The remaining inner-shell
effects are treated with high-level coupled cluster methods such as CCSD(T) or CCSDT(Q).

To further increase the accuracy of our theoretical predictions we have calculated corrections due to
some small physical effects. These include the relativistic corrections (full Breit-Pauli Hamiltonian) and the
leading-order QED corrections. The finite nuclear mass effects (the non-Born-Oppenheimer effects) are found to be
negligible at present.

Spectroscopic parameters generated from the PEC developed in this work show a remarkably good agreement with the
experimental data. This is true for the well-depth (calculated $D_e=934.5\,\pm\,2.5\,\mbox{cm}^{-1}$), dissociation
energy ($D_0=808.0\,\mbox{cm}^{-1}$), and the equilibrium bond length ($R_e=2.4425$ \AA{}). The vibrational energy terms
are on the average
only 1 cm$^{-1}$ away from the empirical results of Merritt et al. \cite{merritt09} showing that the spectroscopic
accuracy has been achieved. We have confirmed the existence of the last (twelfth) vibrational state and predicted that
it lies just 0.5 cm$^{-1}$ below the onset of the continuum. Lastly, this study has proven that the Slater-type
orbitals can routinely be used as a basis set for quantum-chemical calculations for diatomic systems.

\begin{acknowledgement}
The authors thank Bogumi\l~Jeziorski for reading and commenting on the manuscript.
M.L. acknowledges the financial support from the Polish National Science Centre under Grant No. 2016/21/N/ST4/03732. R.M. and J.G.B. were supported by the Polish National Science Centre through Grant No. 2016/21/B/ST4/03877.
Computations presented in this research were carried out with the support of the Interdisciplinary Centre for Mathematical and 
Computational Modelling (ICM) at the University of Warsaw, grant number G59-29.
\end{acknowledgement}

\begin{suppinfo}

The following files are available free of charge.
\begin{itemize}
  \item \texttt{supp.pdf}: contains raw \emph{ab initio} data used for fitting and detailed compositions of the
 STOs basis sets.
\end{itemize}

\end{suppinfo}

\bibliography{ref}

\end{document}